**A ubiquitous ~62-Myr periodic fluctuation superimposed on general trends in fossil biodiversity. I. Documentation**

Adrian L. Melott and Richard K. Bambach

RRH: UBIQUITOUS ~62-Myr periodicity I

**LRH: ADRIAN L. MELOTT AND RICHARD K. BAMBACH**

*Abstract.*—We use Fourier analysis and related techniques to investigate the question of periodicities in fossil biodiversity. These techniques are able to identify cycles superimposed on the long-term trends of the Phanerozoic. We review prior results and analyze data previously reduced and published. Joint time-series analysis of various reductions of the Sepkoski Data, Paleobiology Database, and Fossil Record 2 indicate the same periodicity in biodiversity of marine animals at 62 Myr. We have not found this periodicity in the terrestrial fossil record.  We have found that the signal strength decreases with time because of the accumulation of apparently "resistant" long-lived genera. The existence of a 62-Myr periodicity despite very different treatment of systematic error, particularly sampling-strength biases, in all three major databases strongly argues for its reality in the fossil record.

*Adrian L. Melott. Department of Physics and Astronomy, University of Kansas, Lawrence,*
    *Kansas 66045.  E-mail:* melott@ku.edu

*Richard K. Bambach. Department of Paleobiology, National Museum of Natural History,*
    *Smithsonian Institution, Post Office Box 37012, MRC 121, Washington, D.C. 20013-7012.*
    *E-mail:*  richard.bambach@verizon.net

# Introduction

*The Question of Periodicity in Biodiversity Data.*—There has long been interest in and considerable debate and discussion around the question of periodicities in fossil biodiversity, or often only in the timings of mass extinction. Review of this history is outside the scope of this paper, and can be found elsewhere (e.g., Bambach 2006; Lieberman and Melott 2009). Our work springs from the first high-significance detection of long-term periodicity in fossil biodiversity (Rohde and Muller 2005, hereafter R&M). This relatively recent work is based on marine animal fossil biodiversity over ~500 Myr, using a reduction of the Sepkoski (2002) data (hereafter SEP) combined with the Gradstein et al. (2004) geological timescale. Analysis indicated a significant $62 \pm 3$ Myr periodicity superimposed on the long-term trend. No particular causal mechanism was initially proposed, but the result was published on the basis of its relatively high statistical significance ($p = 0.01$) and potentially strong implications.

*Existing Corroboration of the ~62-Myr Periodicity*.—The R&M work was presented using only one time-series analysis method, and the work was checked and confirmed using alternate methods by Cornette (2007), and Lieberman and Melott (2007); the latter considerably expanded the analysis.

However, these studies were all based on the SEP compendium as reduced by R&M, which had no corrections for systematic errors such as sampling rate. It has been argued for some time that such systematic errors may compromise quantitative study. For this reason, we seek to study this question on the basis of other data sets. An intensive effort (the Paleobiology Database: http://paleodb.org) has resulted in a new  data set, constructed, weighted, and subsampled with the intention of  minimizing such errors (Alroy 2008; Alroy et al. 2008). Melott (2008) extended the analysis to this sample and found evidence of periodicity consistent with frequencies reported initially by R&M.

*Expanded Analysis to Test for Corroboration of ~62-Myr periodicity.*—Here, we further greatly broaden and extend the study of the existence of periodicity in fossil biodiversity in a variety of data sets. In all biodiversity compendia we have substituted the Gradstein et al. (2004) timescale if it was not originally used.

The purpose of this paper is restricted to examining a broadened base of evidence for or against this periodicity in biodiversity. We defer consideration of origination and extinction statistics to future work, and consideration of possible causality to a companion paper. Because large-scale trends in increasing diversity, either from major radiations or from recovery from major extinction events, are driven in part by intervals of elevated origination, and because extinction intensity varies widely from interval to interval, it is not yet possible to extract the subtler variation in origination and extinction responsible for the periodic variation in diversity. A related mathematical point is that derivatives—changes in variables—are inherently noisier than the variables themselves.

The primary periodicity detection of interest to us was based on long-term variation in biodiversity (R&M).  Of course, mass extinctions are part of these trends, but long-term patterns of biodiversity change are what are being analyzed below, not the timing of mass extinctions. Some treatment of spacing in timing of mass extinctions follows in a companion paper. Note that detection of periodicity does not imply that the variation will resemble sine waves.  Nearly any time series can be decomposed into a sum of sinusoids; the question is whether any particular frequencies stand out above the rest.  If they do, it implies at least a partially repeating pattern.

It is important to note that a repeating pattern is only one of many components of biodiversity change. For instance, there is a general trend of increase in diversity throughout the Phanerozoic, with a pause around the Permo-Triassic. The periodic pattern of interest is superimposed on this, and only accounts for a part of overall biodiversity change.  Such signals may or may not be apparent by eye.

In our companion paper to follow, we address relationships of the periodicity demonstrated here to geological data, mass extinctions, and possible causal systems.

## Data Used

In no case have we reduced or altered data, or made any cuts or corrections on data provided us, except as explicitly described in this paragraph.  We will report here on results from the following data: (1) The exact same sequence from SEP called "Well-Resolved Fauna" published by R&M, as provided by Muller, and used in that paper, with no further modifications. Also, two reductions provided by them, called "short-lived genera" and "long-lived genera," i.e., tabulations of those genera that persisted for less than or more than 45 Myr, respectively. (2) Two reductions of the SEP data on genus ranges, prepared by Bambach. One interpolates all ranges to substage level by proportional assignment based on the distribution of ranges known at the substage level. The other restricts the data to standing diversity at interval boundaries, in our case those genera recorded as entering each interval. These data have been illustrated previously with the data available in the supplementary material for Bambach (2006). (3) Reduction of the SEP data by Peters, available at *http://strata.geology.wisc.edu/jack/*. This is a reduction of the data published in Sepkoski (2002) but includes only the genera in which both endpoints (first and last) are known at the substage level. It differs from R&M "well-resolved" in that they interpolated endpoints that were resolved at stage level to the substage. Peters's sorting routine does not interpolate and excludes any genus that does not have both endpoints resolved at the substage level. Therefore this data set is smaller than the others we analyze. (4) The SEP data binned into 106 intervals (a combination of some stages and some substages) using Sepkoski's original sorting routine, which interpolated less well resolved range endpoints among these intervals. Two reductions of this sorting of the SEP data were also analyzed, one combining data on major groups that dominated diversity in the Paleozoic, the other combining data on major groups that dominated diversity in the Mesozoic and Cenozoic. (5) The same Paleobiology Database data published in Alroy et al. (2008), after the sampling rate and other corrections made there, plotted in their Figure 1 and otherwise analyzed there and in Alroy (2008), and provided by Alroy, with no further modifications, hereafter called PBDB.  (6) Summaries of the Fossil Record 2 (hereafter FR2; Benton 1993), marine organisms and continental organisms, downloaded from the website *http://www.fossilrecord.net/fossilrecord/summaries.html*.  In some cases detailed reporting of results has been moved to Supplementary Information, with a summary in this text.

### Fourier Analysis and Its Discontents

Fourier analysis has long been a staple of time series studies in physical science and engineering applications, expanding enormously given the rise of computational science in the last few decades, and the 1965 invention of the Fast Fourier Transform (FFT), and subsequent rise in computational power, making possible extremely rapid evaluation of Fourier series (Cooley and Tukey 1965; Brigham 1988).  Considerable discussion of the modern use of the techniques can be found in Bloomfield (2000), Muller and McDonald (2002) and Press et al. (2007).  The technique is particularly powerful at finding any repeated patterns in series.

The application is based on the fact that nearly any function, given very forgiving conditions, can be decomposed into a sum of sine waves. (To be precise, it can be decomposed into a series of sine waves with phases, or a series of sines and cosines, or a series of complex exponentials; here, we mostly use the first of these options.) These sinusoids span many frequencies. The series begins with the fundamental (the longest that can be fit into the total interval) and accuracy improves with the inclusion of higher and higher frequency sine waves into the sum. In Figure 1, we show how a simple square wave can be represented by a Fourier series. This does not mean that any periodicities we find should look like sinusoids. If the square wave pattern were repeated, that would be easily detectable by examining the series. More examples can be seen in Brigham (1988).

As more and more waves are added, the sum rapidly converges to the desired shape. Normally, many more than the four frequencies represented in Figure 1 are used. With finite time series data, there is a discrete series of frequencies that are used. They are commonly defined by the sampling frequency, the reciprocal of the interval length. The usual upper limit to the frequency is $f_N$, the Nyquist frequency, taken to be half the sampling frequency present in the data. Without delving too much into the large literature on this, we can say that the binning of biodiversity into intervals of time duration $\tau$ is a kind of smoothing procedure that will reduce a problem called aliasing (Press et al. 2007) provided we sample them all, and include frequencies up to $f_N=1/(2\tau)$. Using binning intervals that are too long can introduce problems, as discussed below. We discuss variable sampling intervals below.

Having found the Fourier Transform of a given time series, a wide variety of analytical tools can be used. The autocorrelation (whether computed by FFT or other methods) and power spectrum can be constructed. If we want to compare the time development of more than one series, they can be compared by a cross-spectrum.  Fourier analysis will always find the function expressed as a sum of sinusoids; a common use in periodicity studies is to determine whether any stand out above the rest.  If so, this behavior may or may not be evident in the autocorrelation. There may be severe problems in interpreting periodicity if more than one frequency is prominent.

Note also that we are not studying the precise timing of impulse events such as mass extinctions. Of course the array of extinction events could form a long-duration pattern, which we investigate in a subsequent paper. Certainly they play a role in the lowering of biodiversity, and we will test this role later.  It is the timing of long-duration patterns that is being tested, and hopefully will be clearly illustrated.  Also, we do not require that any cycle we find be the only thing driving biodiversity changes, but merely that it contribute enough to stand out above the background with some statistical significance.

Astronomy and paleontology share the burden of inability to choose where and when one will find experimental results.  Galaxies and fossils are when and where they are found, not on an optimal sampling grid. Geological time intervals are not uniformly spaced.  FFT, as originally formulated, requires sampling at evenly spaced intervals in time or space.  There are two ways to deal with this, and we have used them both as a cross-check.  The first is simple linear interpolation.  This can lead to spectacularly incorrect results, if not understood.  However, if done correctly, it leads to a bandwidth-limited effect, that is a well-understood reduction of amplitude at high frequencies (e.g., see Schulz and Stattegger 1997).  The effect is small when, as here, one is examining time periods longer than the average stage length.  This statement will now be made more precise.

Interpolation is one approach. Simple interpolation has a strong effect on power only at periods close to or shorter than the interpolation (stage) length. We use two methods to determine the amplitude of the various sinusoid frequencies to the function: the familiar FFT in which we interpolate to 1-Myr intervals, and another where no interpolation is needed. Power is commonly defined as the squared amplitude of the contribution at a given frequency. The Fourier Transform of the linear smoothing function (a triangle) is $\mathrm{sinc}^2(\tau\,\pi\,/\,T)$, where $\tau$ is the time window for interpolation, $T$ is the period $(1/f)$ of the frequency f under consideration, and $\mathrm{sinc}(x)$ is defined as $[\sin(x)]/x$. With an 11-Myr bin, the effective sampling window is 22 Myr, as the window is effectively including the midpoint of the two intervals adjacent to one within which the data point lies. The effect of the window is to *reduce* the amplitude of measured frequencies by this factor (squared for the power spectrum). Given the average interval lengths in the reductions, this means that the power of fluctuation would be 1.5 times higher than indicated at the 62-Myr period in the PBDB, with its average 11-Myr bin length and consequent 22 Myr time window, and there would be a much smaller effect in the R&M sample, and others with shorter intervals. The effect of this correction becomes even smaller for longer periods. The effect of interpolation is a well-understood, bandwidth-limited procedure when done correctly. All interpolated results used herein produced "grid" values every 1 Myr, from linear interpolation between the nearest measured values. We emphasize that we do not use the high-frequency information that may result, which would be incorrect.

We can check the results of interpolation and FFT by using an alternate method that is less well known but has the advantage that no interpolation is needed. The Lomb-Scargle Method (Scargle 1982, 1999; Press et al. 2007) was developed in order to use data directly without interpolation, which avoids any amplitude reduction at high frequencies, among other benefits. This is sometimes called Gauss-Vanícek (e.g., Cornette 2007). Its disadvantages include unfamiliarity, and less flexibility in constructing joint statistics between different samples. Also, if one segment of the data is resolved to a shorter interval than the rest, it will introduce results at high frequencies $f > f_N$ based only on this segment; the interpolation method above greatly reduces this problem. As we shall see, both methods agree very well over the frequency range we examine here.

Another procedure is detrending. Failure to remove long-term trends will introduce strong low frequencies, which could mask the oscillations of interest, as discussed in Bloomfield (2000), Muller and McDonald (2002), and Press et al. (2007). For example, a time-series analysis of raw Phanerozoic biodiversity would be dominated by those components needed to represent the general increase in diversity through the Phanerozoic with a pause in the late Paleozoic and earliest Mesozoic. Any oscillation about this overall shape would be hidden. A prime example of this is presented by Omerbashich (2006), who reanalyzed the R&M data without detrending, finding no periodicity. He incorrectly attributed this to his use of Lomb-Scargle/Gauss-Vanícek method (described below). Cornette (2007) and Lieberman and Melott (2007) both verified that regardless of the method of time-series analysis, the periodicity is seen if and only if the detrending takes place first. Although the choice of a best-fit cubic is used in what follows, the appearance of the peak in question is not strongly dependent upon this choice. It is most important to remove the general increase in measured fossil biodiversity between the Cambrian and the Holocene. For more detailed discussion, please see Cornette (2007).

**Results Related to Spectral Analysis of the Paleobiology Database**

*Detrending, Residuals, and Spectral Analysis Procedures.*—We begin with a review of some recently reported results (Melott 2008) that are illustrative of the methods we use for time-series analysis of all the biodiversity data presented here. The methods begin by least-squares fit to a cubic of the new, controlled data kindly provided by J. Alroy (Alroy et al. 2008: Fig. 1). Note that the purpose of fitting (in this case a cubic) is not smoothing, as with a Loess smooth or Savitzky-Golay filter. Smoothing removes high frequencies.  Detrending is rather to remove the long-term trends, leaving fluctuations about those trends intact. For detrending this series, the cubic is the best fit of the various simple alternatives, highlighting the general increase over time, with an inflection possibly associated with long-term effects of the late Carboniferous ice ages and the Permo-Triassic extinction (see Fig. 2A). Its coefficient of determination value is $r^2 = 0.61$, indicating a good fit. A cross-check was done with a linear fit, $r^2 = 0.43$. We also tested the cubic against other possible choices by using the $F$-statistic (Stopher 1975), which weighs the increase in fit of a more complicated equation against the increased number of parameters. The cubic was the best fit by this criterion for all the time series we study here. This cubic fit is less inflected than the fit that R&M show to the data they use. This behavior is a result of the sampling standardization performed in the PBDB, and constitutes part of the new interesting results. PBDB temporal intervals are about 11 Myr, longer than the 3.3 Myr of R&M.

We accept this long-term cubic trend, which is not the main question for us. Our interest is in whether there are any repeating patterns of fluctuation about the long-term trend.   The deviations from the cubic fit of the PBDB data are shown in Figure 2B. For now, these are the data of interest; later we shall discuss the consequences of different detrending choices. The analysis has been done two ways: (1) with Lomb-Scargle based on the data taken as a function of the intervals (and their midpoints), and (2) with FFT on a file constructed by assigning those values to the time of the midpoint of the interval, and then linearly interpolating between them to assign values every 1 Myr.  A time series running 5-520 Ma spaced at 1-Myr intervals was constructed. The period 0-5 Ma was not included in the PBDB sample. Reanalysis of the power spectrum of the data using alternate methods based on the Lomb-Scargle transform, which does not require evenly spaced data and therefore no interpolation, is a robust check of whether interpolation introduced any artifacts. Except for explicitly mentioned exceptions handled by *ab initio* programming, time-series analysis was performed using AutoSignal 1.7. *(http://www.systat.com/products/AutoSignal/).*

This cross-check between Lomb-Scargle and FFT is not really computationally necessary here.  As we describe, it is a well-understood procedure, and the check for equivalence was made in the Sepkoski/R&M data by Cornette (2007) and by Lieberman and Melott (2007). However, some are skeptical of interpolation based on bad history. (For instance, in Solé et al. 1997, results at frequencies close to the inverse of the size of stages were used, so that the $\text{sinc}^2 f$ term greatly affected the results). Others will be skeptical of Lomb-Scargle because it is less familiar than direct Fast Fourier Transform methods.  By showing the results of both analyses, we demonstrate that the periodicity of interest, although slightly different in amplitude, is significant in the results of both procedures.

*Autocorrelation and Memory in the PBDB.*—Herein we summarize results from Melott (2008), with additional description and supporting analyses not included in the original note. This will provide a basis for description of other results based on SEP and FR2. We begin with the autocorrelation function of the detrended PBDB interpolated at 1-Myr bins, which is often used to investigate long-range behavior.  We plot it as a function of time in Figure 3, normalized to its amplitude at zero lag in the detrended data. (Terminology differs between disciplines.

Often this normalized autocorrelation is called the autocorrelation coefficient.) The time series was extended with zeroes, as needed to prevent a spurious "wraparound" effect (often called "zero padding" [e.g., Press et al. 2007]). This figure shows a striking damped-oscillatory pattern, strongly suggestive of repetitive behavior on a period of about 150 Myr.  We note that an analysis with correlation based on lag in intervals without time shows a similar shape. Minor changes are introduced, because the variation in interval length mixes timescales. Alroy (2008) noted that having data correlated with itself at some lag is a necessary condition for periodicity to hold.  However, in that publication, the autocorrelation of biodiversity was plotted out to a lag of ten intervals, corresponding to about 110 Myr. This prevents seeing the full pattern, so Alroy concluded that there was no autocorrelation, which would make periodicity impossible. We see clear indication of periodic behavior as noted above. As we shall see later, there are additional periodicities.  They cannot be seen here, except as alterations in the shape of the dominant curve. A better tool is needed to understand the signals.

*Power Spectra of the PBDB Data.*—Correlation analysis is limited in detecting possible multiple signals, because the value at any particular lag is a sum over all the oscillations in the data, at different frequencies.  In this case we can only see a particular frequency signal clearly because it so dominates. Power spectral analysis is to be preferred, because it separates out the frequencies (Muller and MacDonald 2002; Press et al. 2007).

The power of a given sinusoidal component is the square of its amplitude.  This is directly related to how much it contributes to the variance of the function being studied.  In sound waves and electromagnetic radiation, it tells the energy associated with that frequency.  It has much more tractable qualities for computation than the amplitude.  Formally, the power as a function of frequency (power spectrum) contains the same information as the autocorrelation, but it is segregated by frequency, making it more useful for the purpose of identifying particular frequency components.  There are well-defined procedures for determining the significance as well.  We use those  procedures built into AutoSignal, which in the past have been found (Lieberman and Melott 2007) to agree in detail with Monte Carlo-based results from R&M.

To facilitate clarity, we show the power spectrum in several different ways. The first two are based on the interpolated data as described above, and use conventional Fast Fourier Transform methods with zero padding.  These are the most well known and trusted methods across a wide variety of disciplines. We first show a linear plot of the full power spectrum. Although logarithmic plots are customary for good reasons, they do produce a visual underestimate of the difference in height of various peaks relative to the background, so we will show this first one only on linear and then logarithmic scales. The variance of the residuals is proportional to the area below the power spectrum, and the area under a peak at frequency $f$ is a measure of the contribution of a sine function of frequency $f$ to the variance of the residuals. The variance is clearly dominated by the contribution of a few peaks. The spectrum shown in Figure 4A shows two peaks close to those found previously (R&M; Lieberman and Melott 2007; Cornette 2007), and a new one at higher frequency.  We show a linear plot first, because the log plots that follow may give a false impression that the peaks are small compared with the background power. The lowest-frequency peak has much higher amplitude than seen before. The same data are shown on log-log axes in Figure 4B.  The parallel lines correspond to significance levels of 0.05, 0.01, and 0.001 *against such a peak appearing anywhere in the spectrum.*  We use an auto-regressive (hereafter AR) fit of order one (Press et al. 2007) to the power spectral slope (after filtering out the maxima), corresponding to a "red noise" spectrum (declining power with increasing frequency), as often found in natural time series. Gaussianity of fluctuations around the fit to the spectral slope is assumed, which is common when the fluctuations in a function are the sum of many random variables. We have examined the distribution of residuals of short-lived genera (described later and shown to be the source of the 62-Myr periodicity). They appear approximately Gaussian with some skewness. The skewness parameter is 0.56 and the excess kurtosis parameter is 0.67. Perhaps more meaningful is the comparison of mean (forced to be zero by procedure) and the median 5.7, which is a small difference compared with the standard deviation, 162. This method has been found to agree in detail with results based on the other primary approach, Monte Carlo reshuffling of stages (Lieberman and Melott 2007).

In a recent review, Bailer-Jones (2009) has discussed hypothesis testing, particularly with regard to extinctions. In this, he has discussed (among other issues) the question of testing for periodicity in time series. In view of the discussion there, we feel it is important to clarify for what we are testing, and in particular to reject a number of the "straw man" cases that are presented there. In doing so, it is necessary to digress from our main topic.

We repeatedly examine fluctuation power spectra, and find peaks that rise above a best-fit power law. We examine for peaks that rise above a time series distributed in Gaussian fluctuations around the same power law. If they rise with a probability, say 1% given the Gaussian distribution, we claim significance. Our shorthand for this is rejection of the null hypothesis at $p = 0.01$, and a corresponding "detection of periodicity."

Bailer-Jones (2009) comments that there might be many other null hypotheses. That is, we can imagine a different kind of randomness, which may show such a peak with a different kind of probability. Thus, by rejecting the null model, we are not proving periodicity with 99% confidence.

This is formally correct, if one has some abstract model in mind that is to be "proven." For example, a sine wave is periodic in a strict mathematical sense. One can construct simple devices that have chaotic motion but occasionally show time periods in which they move in something very close to simple harmonic motion (sine waves)—then return to chaotic motion. As a matter of simple experimental fact, one cannot test the hypothesis of periodic motion strictly by any finite time series sample, so the whole thing is a bit of a red herring. The best we can do is to test with the data we have for time sampled—about 500 Myr. Our null model is chosen to represent the most typically seen outcome of time series behavior of systems, which is the sum of many different input influences—a Gaussian power-law spectrum, with the power-law index chosen as a best fit to the data. So our null hypothesis is the complement of the thing tested, and rejection of it implies that the peak is "interesting," in that it rises above what is normally seen. There are no alternatives to "null" or "not-null," and what could be simpler? This means that the time series shows fluctuation at this peak at a level not expected from such a background. Rarely in science do data prove anything—they only make it more probable.

Bailer-Jones uses the example of a deck of cards: Although the probability of drawing the ace of spaces is 1/52, one would not conclude that the deck was rigged upon observing the ace of spades to be drawn from the deck. However, *if the ace of spades had been specified by someone as a prior hypothesis before the draw*, it would certainly be the case that the probability of the deck being rigged would be viewed as having increased by then making such a draw. This is not the place to go into a long discussion of Bayesian statistics, but hopefully the point is taken.

After testing for such spectral peaks we then go on to show that three different compendia of fossil data show signals that are operationally identical, that mass extinctions bear a strong and improbable relation to this signal, and in a paper to follow, that other geological indicators show intrinsically improbable correlations with the same signal.

This analysis was also done (Fig. 4C) with Lomb-Scargle (Scargle 1982, 1989). The existence of the same peaks in Figure 4B,C confirms that interpolation has not strongly affected the results. The AR coefficient fit to the raw, non-detrended, non-interpolated time series is 0.96; significance levels for $p = 0.05$, 0.01, and 0.001 based on this are plotted here. The significance of the peaks is somewhat higher than found with the FFT method. The same agreement of results from different methods was found with different data in earlier work (Lieberman and Melott 2007 and Cornette 2007, which used the alternate nomenclature Gauss-Vanícek). The agreement here between the results of FFT and Lomb-Scargle methods is not quite as precise as with the Sepkoski data (Lieberman and Melott 2007), because the time intervals in the PBDB are longer. Still, the features in question are strong with both methods.

The frequency ($f$) range has been restricted in Figure 4B,C to take account of the limitations due to the length of the total time period and the size of the intervals in the data. The PBDB sample has an average interval length of 11 Myr, and a maximum interval close to 20 Myr. Although Lomb-Scargle may produce results at higher frequencies, based on the Nyquist sampling theorem, reliability degrades around a period determined by twice the average sampling interval $T = 1/f_N \leq 22$ Myr. We have plotted down to $f = 0.04$, or 25 Myr for informational purposes. As discussed before, interpolation windows can artificially suppress power in FFT methods for frequencies near $f_N$ . On the other hand, Lomb-Scargle may show results at and even beyond $f_N$. However, those results at high frequency will result only from sampling the parts of the time series that have shorter periods, and so may not represent any trend present in the data as a whole. These well-understood limitations are complementary.

The major item of interest for us is the peak at $f = 0.0158$. This corresponds to a period of $63.1 \pm 6$ Myr. Error bars presented in this paper are at half the maximum value of the power (which corresponds to about $1.18\sigma$ in the case of Gaussian fluctuations). This peak provides about half the total variance and is significant at $p = 0.01$ (the probability of a peak so rising above the trend anywhere in the spectrum against the background). This is equal within the errors to the Rohde and Muller (2005) result 62 Myr, now at a higher confidence level $p = 0.001$ than when originally found. As we shall see later, we have multiple indications that it is the one least likely to be artifactual and therefore most likely to be biologically interesting. Note that this significance level is against any such peak appearing anywhere in the spectrum. If we had formulated things as whether a peak would appear at the same place, the $p$-value would indicate a much higher significance. However, because we have so many different files to examine, such an approach would quickly become intractable. In addition, the biggest questions about periodicity now involve systematic, not random error, and the real strength of this investigation is finding the same signal in multiple, independent data sets with different kinds of systematic error.

The biggest peak is at the lowest frequency, corresponding to a period of 157 (+24/–20) Myr. It is consistent with a peak detected by R&M at low significance at 140 Myr, but it has much higher amplitude in PBDB. It is a very low frequency peak, with time for only three full periods in the Phanerozoic, but this is naturally taken into account in assigning significance. Note the negative slope of the significance lines, so that longer-period power has to satisfy more stringent criteria for significance. One may be concerned that the three regions in the cubic may somehow contribute to the detection of a signal at one-third the length of the entire interval, as may the three short, sharp diversity spikes at about 400, 260, and 90 Ma in the PBDB (Fig. 2A).

This 157-Myr peak is the component that drives the primary oscillation seen in the autocorrelation in Figure 3. The power spectrum and the autocorrelation contain the same information, but in a completely different way. The 157-Myr power is evident here in the alternating peaks (or alternating troughs) in the autocorrelation, which are about 157 Myr apart. Other spectral peaks contribute to the autocorrelation, but they are swamped by that signal.

Another peak at $T = 46$ Myr rises to high significance, $p < 0.001$. It is, however, not fully resolved from the 63-Myr peak (see Fig. 4B,C). It does not appear at significance in the other analyses done before, or any of the new analyses reported later, all of which have better temporal resolution. We must allow the possibility that these adjacent peaks together represent a single peak, possibly affected by resolution issues. A simple test will support this idea. The interval lengths vary. We have examined their variation by constructing a file of interval lengths in the PBDB data as a function of their midpoints. This reveals a very strong spectral feature at 39 Myr (Fig. 5), which means there is some alternation of shorter and longer intervals with a 39-Myr period. A beat between a 63-Myr signal in fossil biodiversity (for which we shall see there is strong additional evidence) and this 39-Myr sampling variation may have produced a feature at the mean frequency of the two, in this case corresponding to a predicted period of 48 Myr, close to the result found. Future versions of the Paleobiology database with finer temporal resolution will be needed to resolve this issue. In the meantime it seems prudent to regard the peaks at 46 and 63 Myr as one signal, producing about 20% of the variance in the data set. Note that this difference, if our explanation is correct, is a kind of systematic error, not the random error contained in the margin of error.

It should be noted that two peaks close to the first two above actually appear in Alroy (2008, Supporting Information, Fig. S2C), and rise substantially above the 95% confidence lines.

Lines for higher confidence levels are not plotted there, but one may infer that the peaks would surpass them. They are described as insignificant, but the data suggest otherwise. For this reason it is desirable to introduce additional statistical tests, across multiple data sets. This multiplicity is the primary motivation of this paper.

*Cross-Spectral Analysis of PBDB with Sepkoski/Rhode and Muller Data.*—Additional testing of the objective reality of features in fossil biodiversity may be gained by doing statistical tests that combine data from different, independently compiled data sets.  The SEP data used in R&M and the PBDB sample-standardized data we have used here thus far were collected and treated very differently.  We have seen that two signals at similar frequencies arise in both; these two are within the errors on their frequencies, and so are good candidates for further investigation.  It is valuable to test these features for combined amplitude.

We now describe a combined analysis, additionally using data as downloaded from Supplementary Information in R&M, specifically the "Well-Resolved Genera" sample emphasized in R&M as well as Cornette (2007) and Lieberman and Melott (2007).  This is done by analysis of the cross-spectrum (Press et al. 2007) of the two detrended series. It is most easily described using the complex exponential representation of Fourier series. This is a generalization of the power spectrum of a single time series, which is essentially $A_i*A_i$, where * denotes complex conjugation and $A_i$ denotes elements of a series of complex Fourier coefficients as a function of frequency.  The power spectrum is therefore real-valued.  The cross-spectrum involves the Fourier coefficients of two different series: $B_i*C_i$, and it is complex.  The amplitude of this complex number is a measure of the extent to which a given frequency is present in both series; its phase is a measure of the extent to which the components of the two series are in phase, i.e., whether the timing of the peaks and troughs coincide.  To the extent that the signals are anchored in objective reality, we expect phase agreement between components that came from different data sets.  If they are in perfect phase, the cross-spectral coefficient is real and positive near the peak.  Out-of-phase signals with the same frequency will not be positive real numbers—they will be complex numbers, and may have a negative real part or have a substantial imaginary component. Thus the test for a positive peak in the real part of the cross-spectrum is a stringent test that requires not only that both data sets have the same frequency at strength but also that the signals have some component in phase—peaks and troughs coincide in time. We can also examine the imaginary component, as done later, and see whether it is comparable to the real positive part, which would be further indication that they are out of phase. We have computed this directly, Melott writing our own software, supplemented by IMSL Fast Fourier Transforms, as AutoSignal does not have cross-spectral capability. The time period 5-505 Ma, common to both data sets, is used. In order to give the data sets equal weight, each has been divided by its own standard deviation after detrending and interpolation to 1 Myr.

In Figure 6, we show Real($C_{sp}$), the real part of the cross-spectrum, as a function of frequency.  Peaks found in Figure 4 are also seen here.  Peaks at 156 and 47 Myr are lower than the total amplitude of the cross-spectrum would indicate.  This is because, at these places, the function has substantial imaginary component. Components in the two data sets are out of phase by 1.34 radians (or 33 Myr) at 156 Myr, and by 0.68 radians (or 5 Myr) at 47 Myr. Consequently the objective origin of these is called into question, because although both periodicities appear in the cross-spectrum, this tells us the peaks and valleys are not at the same times in the two data sets. We do not have cross-confirmation of these peaks because of this mismatch.  If they originated in actual changes in biodiversity as found in both data sets, the phases should have good agreement. They may be affected by boundary conditions in one case and temporal resolution in the other.  We will conduct some further tests.

On the other hand, the cross-spectral peak at 61.7 (+4/–3) Myr is completely robust.  Its phase displacement is only 0.16 radians between the two very different data sets, corresponding to a 1.6-Myr difference in the placement of peaks and valleys of a much longer cycle. This strong agreement suggests that further discussion should be about possible causes for this signal, rather than its reality.  It is robust against choice of interval length and data selection procedures, detrending algorithm and analysis procedure, with strong agreement between the two as found in the cross-spectrum.  Note that the phase agreement is very strong, and in fact the difference between the two is much smaller than the errors on either measurement.  This strong result needs to be further tested, which is a central goal of this paper.

*Alternate Detrending and the ~150* Myr *Signal.*—Detrending was conducted using a cubic after previous work (Rohde & Muller 2005; Cornette 2007; Lieberman and Melott 2007), because a cubic is such a good fit both by eye and the best fit of many alternatives by the objective measure of the Coefficient of Determination, adjusted for degrees of freedom. Nevertheless, there may be a problem here.  First, this longest-period signal appears at different levels in those studies versus the results shown here in Figure 4A–C, where it is much stronger, and in fact dominates.  Second, our cross-spectrum test shows that the peaks of this longest-period cycle are strongly misaligned between the two data sets.  Methodologically, one may note that the cubic "rounded Z shape" has three broad regions, and may somehow contribute power at approximately one-third of ~500 Myr, the total interval length.  For all these reasons, it is desirable to do additional tests of sensitivity to the detrending procedure.   We will show one example in some detail, then report briefly on alternatives.

All that is required for time-series analysis is removal of long-term trends. The most basic detrending example is linear. In Figure 7A, we show the result of a least-squares fit to a straight line, and in Figure 7B, power spectrum of the residuals resulting from this choice. Because some peaks are reduced in amplitude, we have added lines at 50% and 90% confidence. The suspicious long-wave signal at 157 Myr has dropped far below significance, not even reaching $p = 0.5$. The other two survive, peaking at 63 Myr and 48 Myr. (Note that because the linear fit deviates from the straight line in opposite directions at the ends, a tapered-cosine window has been applied. The only effect of this function is to force the function to zero at the ends, with no effects elsewhere. Actually, though formally desirable for the use of FFT methods, we found that it made no practical difference in the results.

We performed a separate test to test whether the three high peaks, exceeding 150 in Figure 2B, may be the origin of the enhanced 157-Myr signal in the PBDB. This was done by truncating these peaks and two valleys at an amplitude of 80, typical of the amplitude of other excursions in these detrended data. This change did not result in much reduction of the low-frequency spectral peak. Consequently, we conclude that these swings of very large amplitude are not the primary cause of the low-frequency feature that is very strong in this data set.

We have also explored the PBDB power spectrum with detrending by functions that are best-fit quadratic, logarithmic, exponential, power-law, and hyperbolic functions. In all these cases the peaks near 62 and 46 Myr retained their significance at about the $p = 0.01$ level or better, but the 157-Myr peak ranged from $p = 0.1$ to $p = 0.01$, depending upon the function. This dependence on the shape of the detrending signal is troubling. All these functions have low-frequency changes in slope. This behavior, along with the fact that cubic detrending gives a signal about 13 Myr out of phase between our two primary data sets, causes us to question its reality, and in particular that it may be related to the difference in shape of the overall cubic fits to the two data sets. On the other hand, the other two signals survived all these permutations.  In fact, the 62-Myr peak is strongly visible even with no detrending.

**Results Based on Various Reductions of the Sepkoski (2002) Data**

*Summary of Previously Published Analyses.*— In this section we will examine the behavior of the periodicity of interest in a variety of reductions of the SEP data set.  Because the results are generally similar, we summarize several of them, putting the detailed results in the supplementary material (http://dx.doi.org/10.1666/09054.s1). Insights gained from examination of short-lived versus long-lived genera are discussed below. Because much of the published work on periodicity in diversity to date is based on the Rhode and Muller reduction of SEP, we summarize those results first.

We briefly discuss, but do not repeat in detail, analyses of Rohde and Muller (2005), who found a $p = 0.01$ power spectral peak at 62 Myr in their reduction of well-resolved fauna from the Sepkoski (2002) data with dates from Gradstein et al. (2004). This work was independently checked by Cornette (2007) and Lieberman and Melott (2007; hereafter L&M) with no ensuing disagreements. R&M used interpolation and FFT, but this was avoided in both follow-up studies by the use of Lomb-Scargle methods.  All results were essentially identical.  L&M did a number of additional detailed analyses, of which only a small part will be emphasized here: (1) The 62-Myr peak is stronger when examining fractional biodiversity changes (changes divided by the total number of genera) rather than absolute ones. (2) The signal weakens for times later than 150 Ma, and is absent for long-lived fauna, i.e., those genera that survived more than 45 Myr. (3) The signal is much stronger for short-lived fauna, i.e., those genera that survived less than 45 Myr.  We view this difference as interesting because, although we would expect the 62-Myr fluctuations in diversity to be magnified in the short-lived genera, it is not obvious why the signal should be missing from the diversity history of long-lived genera. (4) The 62-Myr peak cannot be attributed to just extinction or just origination rates alone; they interact and both must contribute to the signal. (5) Eliminating any one of the three largest extinction events from the time series does not remove the 62-Myr spectral peak. (6) There is a feature at 27 Myr in the power spectrum of interval lengths as used in the R&M sample, that is, all substages, in which ages of stage boundaries are derived from the Gradstein et al. (2004) timescale and substage boundaries are assumed to be equidistant between stage boundaries. Because major and minor extinction events are commonly used to designate stratigraphic boundaries, this feature may be related to the old claim of periodicity in peaks of extinction (Raup and Sepkoski 1984, 1986), but, with recent improvements in the dating of the time scale, now appears to run through the whole Phanerozoic. However, there is no feature in the power spectrum of interval lengths at or near 62 Myr.

*Sepkoski-Bambach Substage Data*.—This is a complete compilation of all marine genera compiled by Sepkoski as of 1996. The publication of 2002 was posthumous and is derived from the files he had in 1997–98. Sepkoski gave this data set to Bambach in late 1996, and it varies from the final by only a few hundred genera out of over 35,000. Bambach worked to interpolate all the data to the substage level. Sepkoski's own sorting routine did not use all substages, but this version does.

This is the only way we have of comparing the R&M data (which omit some genera, both those with less well resolved endpoints and singletons resolved only to stage, which is why their numbers are smaller) directly with Sepkoski's full data set. Also, this is the complete genus diversity at the substage level, so in a way it is the standard from which all other Sepkoski-based data are derived. This contains 165 substages, and thus combined with the substage diversity information, has the best time resolution of any of our samples. We remove data before 518 Ma, because the numbers are very small, leaving 158 stages. The computational result, seen in Figure 8, shows the existence of a peak at $62.4 \pm 3.3$ Myr, at a confidence level better than .001, which higher significance cannot be measured by our software.

*Analysis of Additional Reductions of the SEP Data*.—We additionally performed spectral analysis of four other reductions of the Sepkoski data.  More-detailed information on these results is present in the Supplementary Information for this paper.  These four reductions are (1) a standing diversity estimate prepared by Bambach, which is effectively an estimate of the diversity at each interval boundary; (2) the full Sepkoski data set sorted by Sepkoski's own sorting routine, based on 107 intervals; (3) Peters's "all well resolved" reduction, which includes only genera with both ends of their ranges resolved to the substage level; and (4) Peters's "multi-interval well-resolved" reduction, which is like (3) but with the additional restriction that genera must cross at least one substage boundary—no singletons.

We briefly summarize these results by saying that the periodicity near 62 Myr originally found by R&M appears in all reductions of SEP. They all show a significant periodicity with a period ranging from 61.7 to 63.3 Myr. The phase angles are all similar, as shown from the cross spectra, showing a general coincidence of peak and trough times. The significance levels vary considerably.  For (1) and (2), they are better than 0.001 and exceed our ability to measure them with AutoSignal software. The peak in (3) is the least significant, with $p < 0.05$, but the significance recovers somewhat for (4) at $p < 0.01$. For details see Supplementary Materials, Section I and Figure S1A–D.

## Issues Related to Short-Lived and Long-Lived Genera

*Introduction and Exploration of Dominants from Selected Eras.*—R&M, as well as L&M (their Figure 3), showed that long-lived genera (defined as those genera that survived more than 45 Myr) do not show significant periodicities, whereas short-lived genera show them strongly. Both R&M and L&M found that the overall periodicity is derived from short-lived genera (defined as those that survived less than 45 Myr). Although any single choice for categorizing short and long-lived genera might seem arbitrary, 45 million years is the mean length of the periods of the Phanerozoic segment of the geologic timescale, so parsing genera into those that survive no more than the average length of a period in the timescale and those that survive for more than the average length of a period and would always be found in several major time intervals is a geologically reasonable choice.

However, there is also an apparent time-dependence of the signal. Specifically, L&M showed (their Figure 4) that the 62-Myr periodicity did not exist at a significant level in the last 150 Myr. Because the signal is strong throughout the Paleozoic, but appears to fade in the later Mesozoic and Cenozoic, we checked to see if there is any obvious difference between Paleozoic and Mesozoic-Cenozoic faunal dominants that might contribute to the change in the signal. Higher taxa that diversified later in time might have biological characteristics that are relevant to their surviving whatever causes the downturns in biodiversity that characterize the Paleozoic and early Mesozoic. If there were a systematic difference in higher taxa but not in the background process, we might expect the periodic signal to persist into the last 150 Myr in the group of taxa that had reached their maximum diversity in the Paleozoic but not in the taxa that came to dominate diversity in the Mesozoic and Cenozoic. If the periodicity is present throughout the Phanerozoic in both Paleozoic and Mesozoic-Cenozoic diversity dominant groupings, groupings, then the long-lived versus short-lived difference should be related solely to the longevity of the genera in the two longevity categories and not a preferential bias rooted in biological differences between the suite of taxa with maximum diversity in the faunal dominants of the Mesozoic-Cenozoic or the Paleozoic.

We found that the signal of interest was *present in both groups*, but weakened for recent times. For details see Supplementary Material, Section II and Figure S2A,B. We now further examine the short versus long-lived question.

*Time-Series Analysis of Short-Lived and Long-Lived Genera*.—L&M did not examine the recent behavior of the short-lived genera separately.  In Figure 9A we show the power spectrum for the period 45-295 Ma for short-lived genera. The nature of the sample obviously requires starting at 45 Ma. The data are admittedly noisy, and the shorter length of the interval increases the level of background fluctuation (a problem we may have seen with the ~150-Myr fluctuations in the overall data set). Two new features appear for this interval only at 46 and 38 Myr; we will discuss them later.  But importantly, the peak seen before arises at high amplitude at about 65 Myr (Figure 9A) and, importantly, with a phase angle consistent with its appearances in the three analyses of the full data set. (We note in passing that an exponential was best-fit by *F*-statistic and used for this detrending, but that use of linear detrending still gives a significant peak).  *So, there is no fading away of the signal for short-lived genera at "recent" times.  And it was never there at a significant level for the long-lived genera.* (We also note that a spectral peak near 62 Myr exists for the long-lived genera, and again with a consistent phase angle, but it is not significant, i.e., does not stand up above the level of other components).

How then can we explain the fading of the overall signal, mirrored in its reduction in the subsamples of the previous sections?  The overall signal is composed of the sum of the short-lived and long-lived fauna.  The proportion of long-lived fauna changes with time, as more and more "survivors" accumulate. In Figure 9B we show the proportion of short-lived genera as a function of time. There is a dramatic reduction throughout the Phanerozoic. Particularly after 150 Ma the value remains quite low. Therefore, after 150 Ma the short-lived genera contribute very little to the signal derived from the total fauna. The signal has become diluted by the sheer number of accumulating long-lived genera that vary around their trend line, but not with the periodicity shown by short-lived genera. Further analysis of this issue, including patterns of origination and extinction, is presented in a forthcoming paper.

In a classic study of Ordovician cohorts of genera, Miller (1997) demonstrated that it takes time for species in a newly evolved genus to diversify, it takes time for new species to evolve that are adapted to different environments than the initial species founding a genus, and it takes time for species in a genus to become geographically widely distributed. Jablonski (1986, 1989) has also demonstrated that taxon richness, widespread geographic range, and broad environmental tolerance confer extinction resistance in "background" times. Therefore, we should expect that recently evolved genera would be more vulnerable to periodic increase in stresses than well-established, diverse, widespread, and environmentally tolerant genera. As long-lived genera accumulated in the global marine fauna, their increased numbers eventually swamped the signal of periodicity that characterizes the diversity history of more-vulnerable short-lived genera. One might speculate that this is a "natural selection effect": survivors of whatever stresses that drive the biodiversity cycles expressed by the fluctuation in diversity of short-lived genera would have more time to diversify taxonomically, ecologically, and geographically, and may have become better able to survive the next round. That would have led both to the likely increase in the number of genera that would become long-lived genera and to a weakening of the effect of the repeating stress system even on newly evolved genera, although they still would be generally more vulnerable to extinction and so would still frequently end up as short-lived genera.

## Results Based on Fossil Record 2 Data

*Marine Biodiversity.*—Further independent testing of the hypothesis of a 62-Myr periodicity in marine diversity can be gained from looking at the FR2 data (Benton 1993, 1995), in particular, existing summaries downloaded from *www.fossilrecord.net/fossilrecord/summaries.html*.  In this case, however, we have updated the data by substituting midpoints of stages computed from Gradstein et al. (2004) dates in place of the ones used in FR2.  This is a somewhat challenging test, because FR2 is only resolved to the level of families.  This consolidation compared with the other data sets will reduce both the signal and the noise, with unpredictable results.  On the other hand, FR2 presents certain opportunities, due to different partitioning of data and the inclusion of plants and terrestrial organisms. If the periodicity is present in the marine data from FR2 then we have the opportunity to see if the same periodicity is present in the terrestrial record as well. Recent work suggests that the family-level sampling of continental organisms is reasonably representative (Kalmar and Currie 2010).

We begin with the summary closest to that which we have examined before, which is the "minimum" number of families of marine organisms. Minimum implies the most conservative (i.e., best documented and least ambiguous as to environmental relationship) choices of families to be included in a given interval (Benton 1995). Using the same methods described for previous analyses, we show the (Lomb-Scargle based) power spectrum for marine families (minimum) from FR2 in Figure 10A.  There is a spectral peak corresponding to a period of $61.3 \pm 2.5$ Myr, $p \sim 0.001$.  When done with FFT analysis (not shown) this peak moves to 61.1 Myr.  We may test this for consistency by doing a cross-spectrum as before with the R&M data, with the result shown in Figure 10B. Once again there is a peak corresponding to the same period, within errors; the agreement is closer than the width of the peak in either compendium.  This additional test provides the advantage that it includes all families of organisms preserved in the fossil record, whereas previous data sets included only animals.  We have also done the FR2 marine families (maximum) and found a peak in the same place, $p = 0.01$.

We have so far done cross-spectra of each of two data sets, the PBDB and the FR2 marine families, with the R&M data.  A complete comparison will include cross-spectra of FR2 and PBDB with one another.  This is shown in Figure 10C.  Once again a peak close to 62 Myr appears as in significant agreement between the two data sets, this time showing an average offset of only 0.57 Myr, corresponding to an extremely small difference in phase angle. The lower-frequency peak emerges at 156 Myr, but with a significant phase disagreement of 18 Myr between the two data sets.  No other spectral peaks appear with any significant agreement between the two samples.  Thus, there is a consensus between all three samples, despite their differences in temporal resolution, interval boundaries, and methods of culling data: one peak near 62 Myr shows excellent agreement among all three, and another peak around 150 Myr emerges in all three, but has considerable phase disagreement and dependence upon detrending procedures.

*Fossil Record 2 Terrestrial Biodiversity Data*.—Additional information, useful for eventual investigation into the causes of fluctuation, can be gained by examining continental families, available from the FR2, but not available in data restricted to marine organisms only. Because a significant terrestrial flora and fauna do not exist until the Silurian, this is truncated at 420 Ma. The easiest way to see the essential result is a spectral plot (Fig. 10D).  None of the peaks are significant relative to the background. If the "maximum," i.e., a broadly inclusive, FR2 continental summary is used, a peak does appear at 95% confidence, period 59 Myr, phase generally consistent with the others.  However, the "maximum" set includes equivocal and shared habitat designations (i. e., families that have marine as well as terrestrial members and families with poorly known habitat preferences) (Benton 1995), so we conclude that there is no reliable terrestrial periodicity signal in the FR2 sample.

There are two rather obvious possible explanations for this: (1) Whatever is causing the signal for marine organisms is not affecting terrestrial family diversity; or (2) the small sample size and reduced time interval with good data is masking the signal. Because the signal is clear in the marine record for the interval represented in the terrestrial record, and is even present in shorter sections of the marine record comparable in age with the later portions of the terrestrial record, we suggest that the signal is simply absent from the terrestrial record, at least at the family level.  This conclusion is consistent with the work of Kalmar and Currie (2010).

### Summary of Results and Their Interpretation

We have demonstrated the existence of a periodicity around 62 Myr, which is superimposed on other changes in fossil biodiversity.  Its existence in multiple data sets and different treatments of the same data argue for its objective reality, independent of details of collection and analysis. It has appeared with two independent approaches to Fourier analysis, using FFT plus interpolation, and Lomb-Scargle analysis.  It has appeared in many different reductions of the Sepkoski data set with differing levels of culling and with varying intervals lengths, whose number ranges from 107 to 158 (after truncation)  in the Phanerozoic. It appears in the Paleobiology Database with its strong sampling-bias corrections, and in the Fossil Record 2 marine families, with 79 intervals, but always at very nearly the same frequency, with a phase such that the peaks and valleys closely coincide. The fraction of genera resistant to whatever is causing it increases with time.

In examining the evidence for and against extraterrestrial causes for climate change and mass extinctions, Bailer-Jones (2009) discusses the many long-period signals that have been claimed over the years.  He concludes that the 62-Myr period has the best support of any.  He raises a number of questions, nearly all of which we have answered.  We have earlier commented rather briefly in reply to his considerable discussion of hypothesis testing, but will illustrate here how this study answers nearly all of the further issues he mentions.

He comments that it is suspicious that the signal seems to vanish for data more recent than about 150 Ma, as pointed out by Lieberman and Melott (2007). We now can actually follow the periodicity in the short-lived genera beyond the 150 Myr point (and we have already shown that the loss of signal is because of the rise to diversity dominance of the long-lived genera).

We also have the fact that the periodicity is located in one set of data—the short-lived marine genera—and is not some "artifact" of methodology. If it were showing up in all data (long-lived genera and terrestrial FR2 data) then complaints about methodology might be a possibility. But because the long-lived marine genera and the terrestrial families do not show the periodicity but the marine families and the short-lived marine genera do, we know that it is a phenomenon associated with a particular set of data, not something superimposed on any data treated with our methodologies. Conversely, since it appears in more than one set of data, with more than one method of generating power spectra, it is unlikely to be those methodologies generating a false signal.

Bailer-Jones also asks about the sensitivity of the signal to dating inaccuracies.  It is true that we have used only the Gradstein et al. (2004) dates (except for the Exxon sea levels, where substitution is not possible).  We are not interested in substituting an older, presumably less accurate timescale built on less information.  However, we would argue that the presence of the signal in many different partitions of geologic time with different numbers of stages used argues for insensitivity to this problem.

He argues that the 62-Myr signal may be a fluctuation in the efficiency of fossilization, rather than in biodiversity.  We address this issue in another paper.

## Summary

1.  We have demonstrated and explained that the 62-Myr periodicity is real—it is present in the three different data sources for global marine diversity (SEP, PBDB, FR2).

2.  We pointed out that the periodicity is superimposed on the larger-scale trends in marine diversity (the three radiations and the diversity drops and recoveries from the "big-five" mass extinctions), so it is a curiously pervasive phenomenon, although not the main driver in the history of life.

3.  We demonstrated that the periodicity is a marine phenomenon. It is not visible in the FR2 non-marine data.

4.  We extended previous results that show that the periodicity is expressed strongly in the short-lived fauna, and undetected in the long-lived.

5.  We pointed out that because the periodicity fades out as the proportion of long-lived genera increases ("swamping out" the periodicity that is expressed preponderantly by short-lived genera), it appears that the world's fauna has actually evolved some "extinction resistance," something others (e.g., Stanley 2007) have suggested would accompany the declining rates of origination and extinction that characterize evolutionary dynamics through time.

Now we need a cause for the observed phenomenon of periodicity.  We will look at some clues in a companion paper.

## Acknowledgments

We thank J. Alroy, M. Benton, and S. Peters for providing some of the data which we analyze herein. Two referees provided extremely helpful suggestions on presentation, and Don Johnson provided the illustration for Figure 1. We are grateful to the American Astronomical Society for the sponsorship of the 2007 Honolulu multidisciplinary Splinter Meeting at which discussions leading to this project took place. Research support at the University of Kansas was provided by the NASA Program Astrobiology: Exobiology and Evolutionary Biology, under grant number NNX09AM85G.

**Figure Captions**

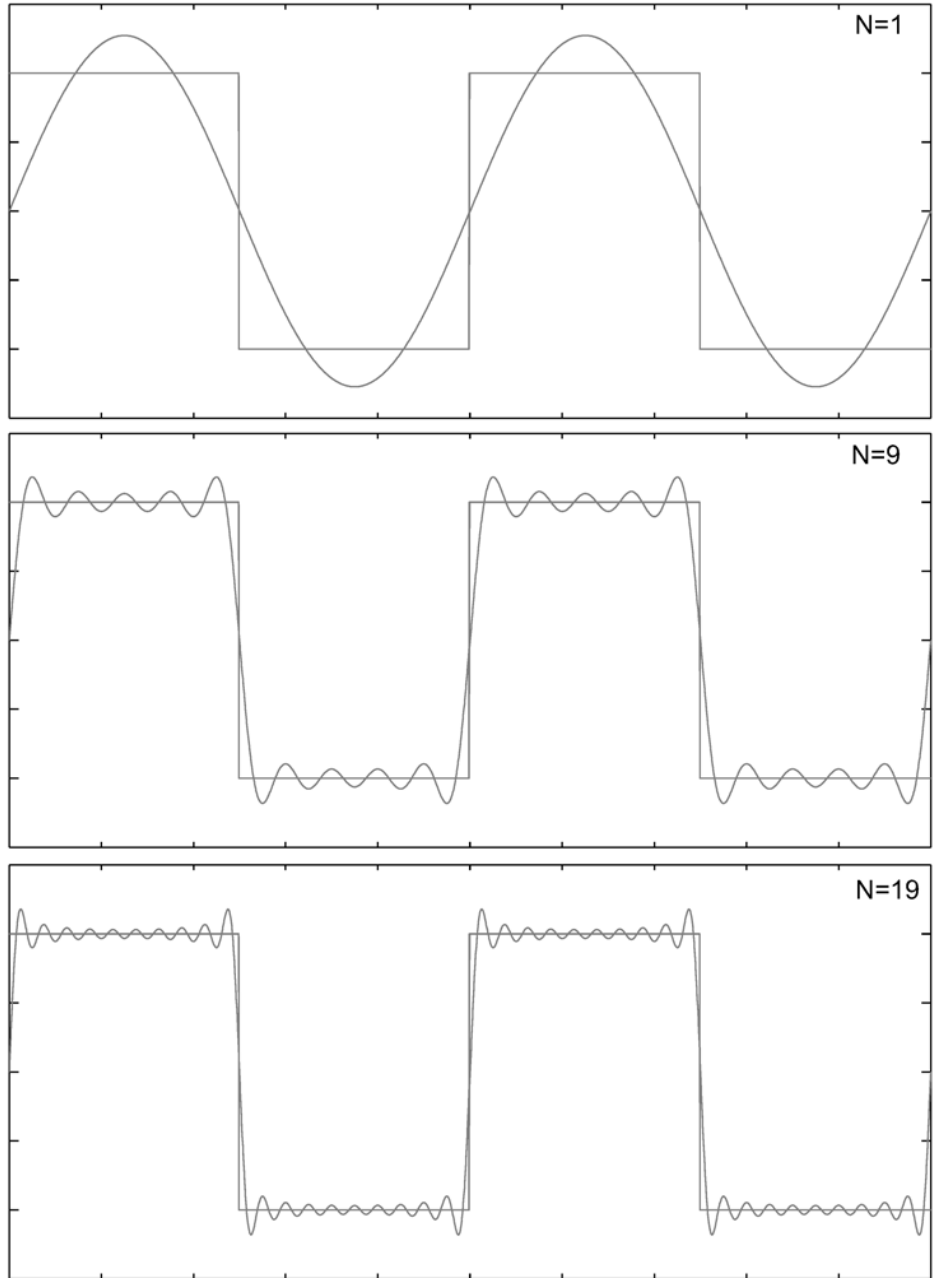

FIGURE 1.  This panel shows how a square wave shape may be represented as a sum of sine waves.  The convergence to the target shape is quite rapid, and any desired degree of accuracy may be achieved by adding more terms to the series.  Illustration provided by Don Johnson.

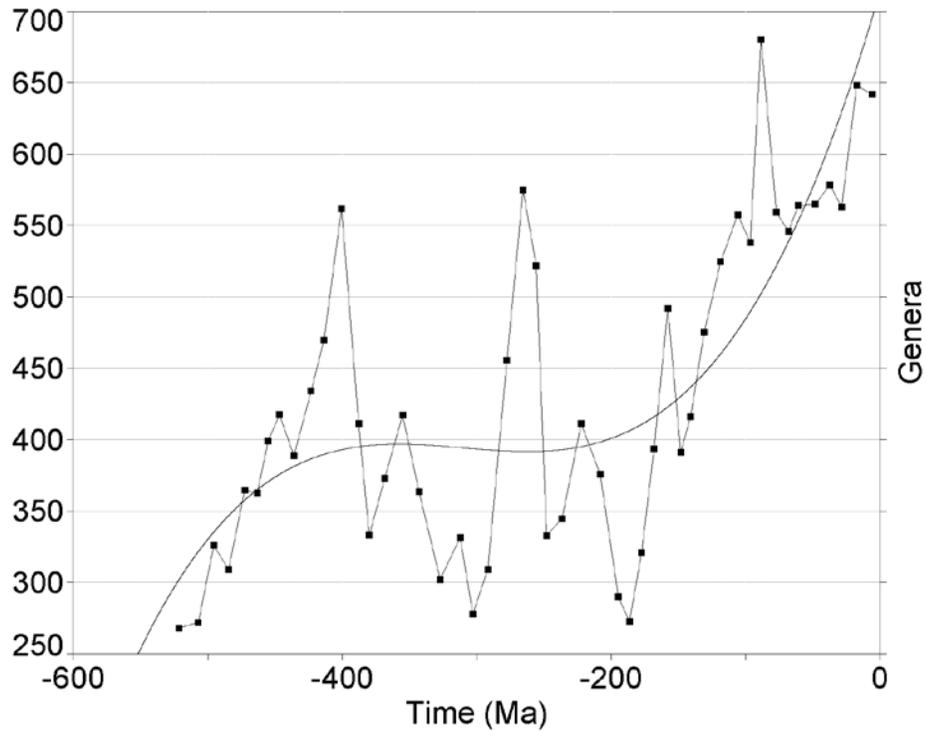

FIGURE 2. A, The sample-standardized number of Genera from PBDB plotted against the date of interval midpoints as shown in Alroy et al. (2008: Fig. 1), along with a cubic fit by least-squares. The purpose of the cubic is to remove the long-term trend, so that we may study fluctuations about this trend.

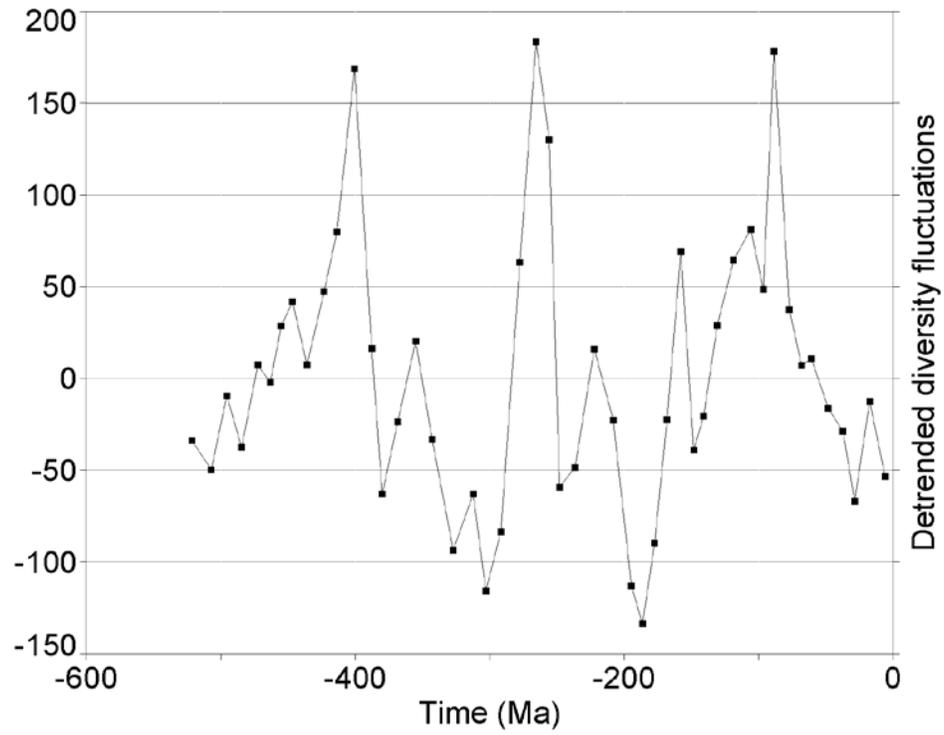

2B, The residuals of the genera from A, differenced from the cubic fit.

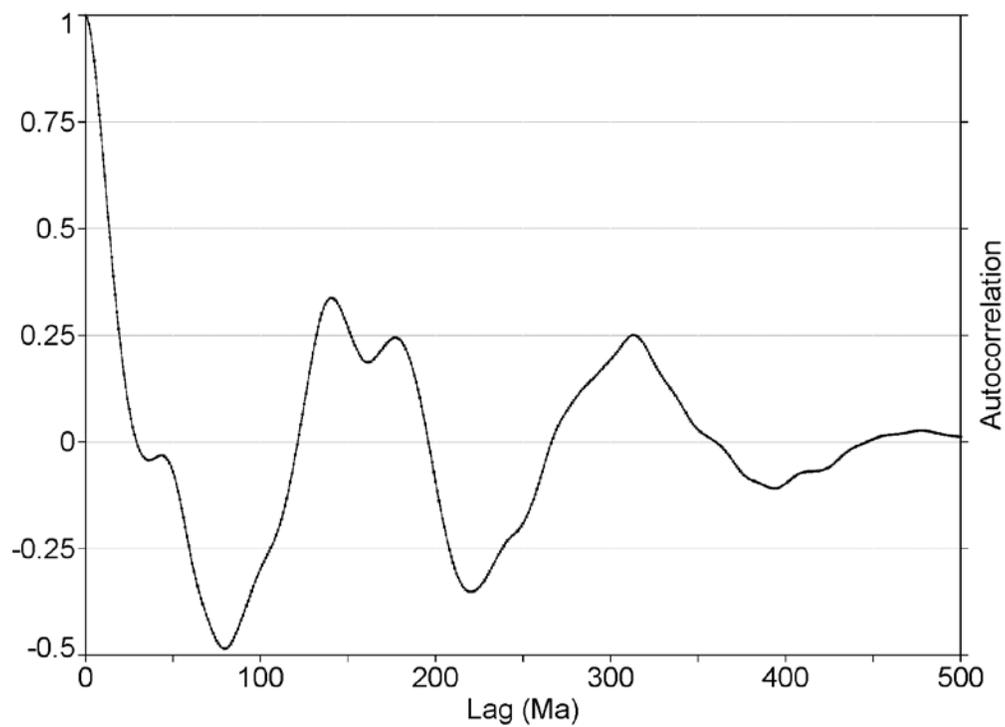

FIGURE 3. The autocorrelation (Melott 2008) of detrended, sampling-standardized fossil biodiversity in PBDB, normalized against its value at zero lag, as a function of time. Note that there is an alternating pattern of peaks and troughs with a period of about 150 Myr, and extending with declining amplitude to the entire sample interval.

*doi:10.1371/journal.pone.0004044.g001*

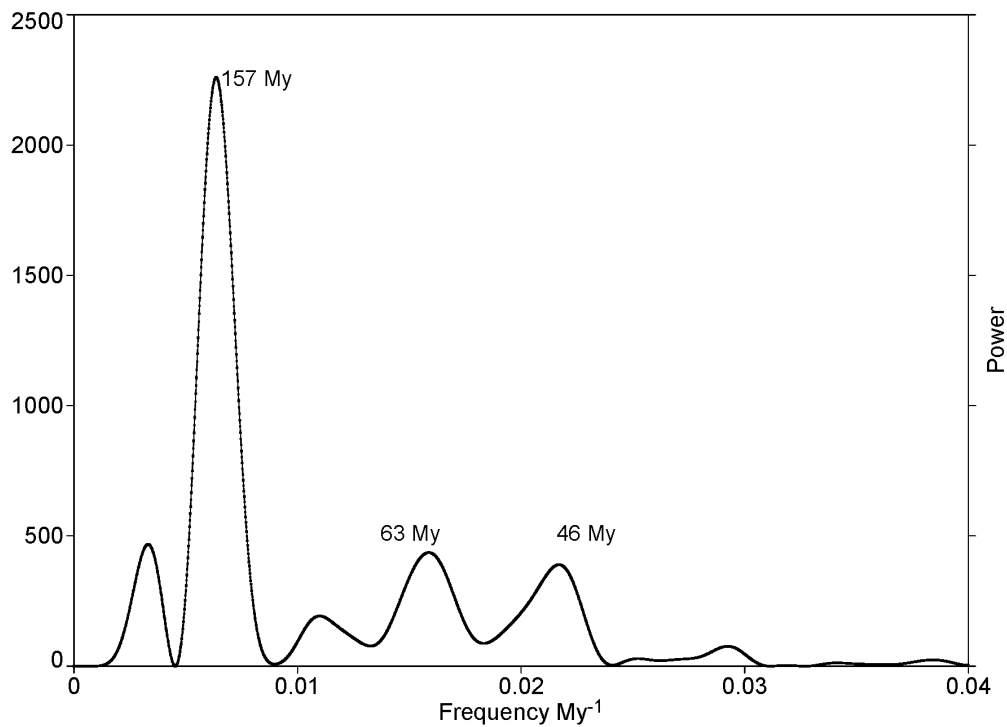

Figure 4. A, A linear plot of the power spectrum of fluctuations in PBDB (Fig. 2B) (determined by Fast Fourier Transform) against frequency in Myr$^{-1}$.  Higher-frequency fluctuations are not shown due to sampling limitations (too close to the interval time scale). The total power is dominated by the area under a few high peaks.  These peaks are labeled with the period $T = 1/f$ corresponding to frequency at the peak.

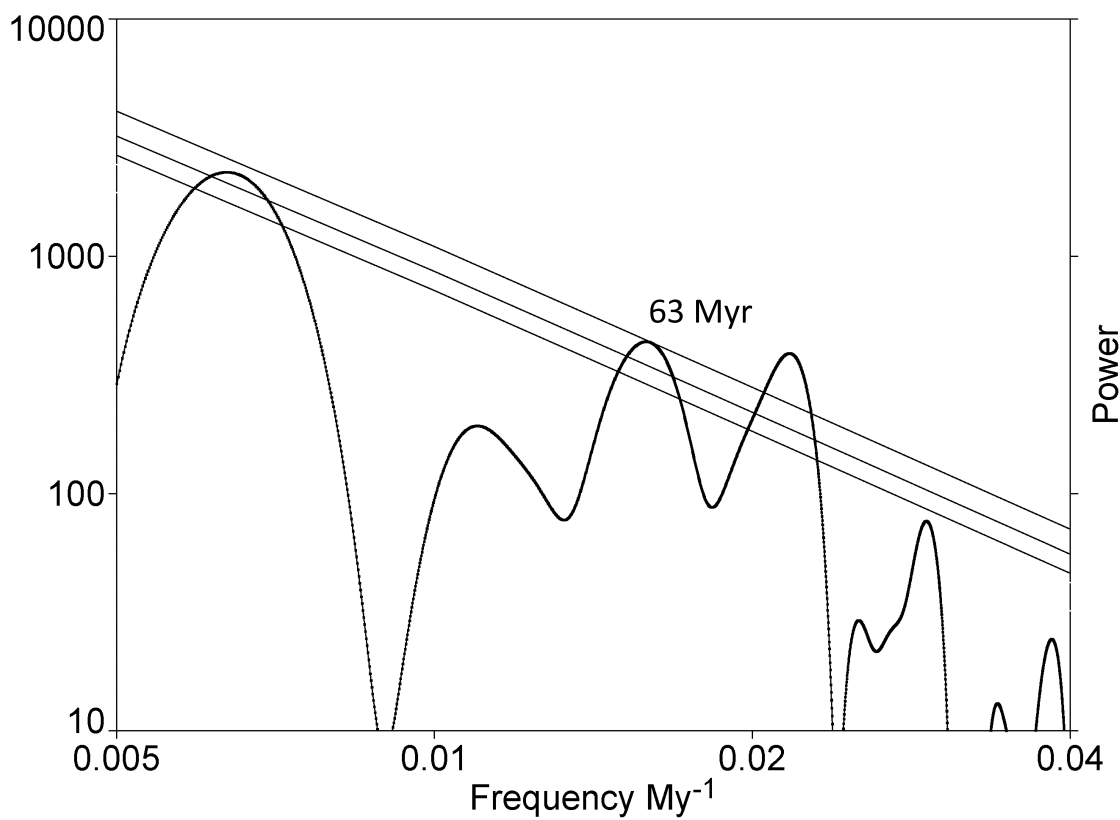

4B, Logarithmic plot of the power spectrum of fluctuations in PBDB (Melott 2008) as in A

(determined by Fast Fourier Transform) against frequency in Myr⁻¹. Fluctuations outside this frequency range are not shown due to sample limitations (interval length and overall time range).

The parallel lines indicate significance at levels $p = 0.05$, 0.01, and 0.001 against the probability of any such peak arising against the spectral background.

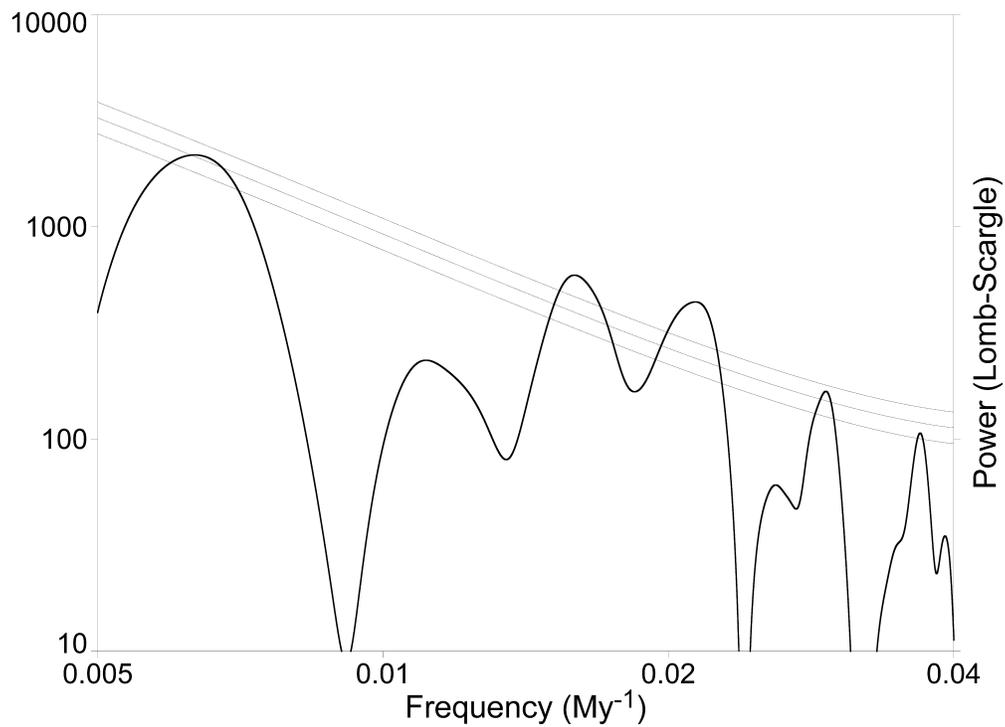

4C,  Logarithmic plot of the Lomb-Scargle transform based power spectrum of fluctuations in

PBDB.  Peaks close to those determined by FFT appear in this plot.  Amplitude at the 62-Myr peak is reduced in B as compared with this result,  marginally visible on this log plot, and less so at longer periods.  Thus the two methods agree very well for periods long compared with the interval length, as would be expected from elementary principles.

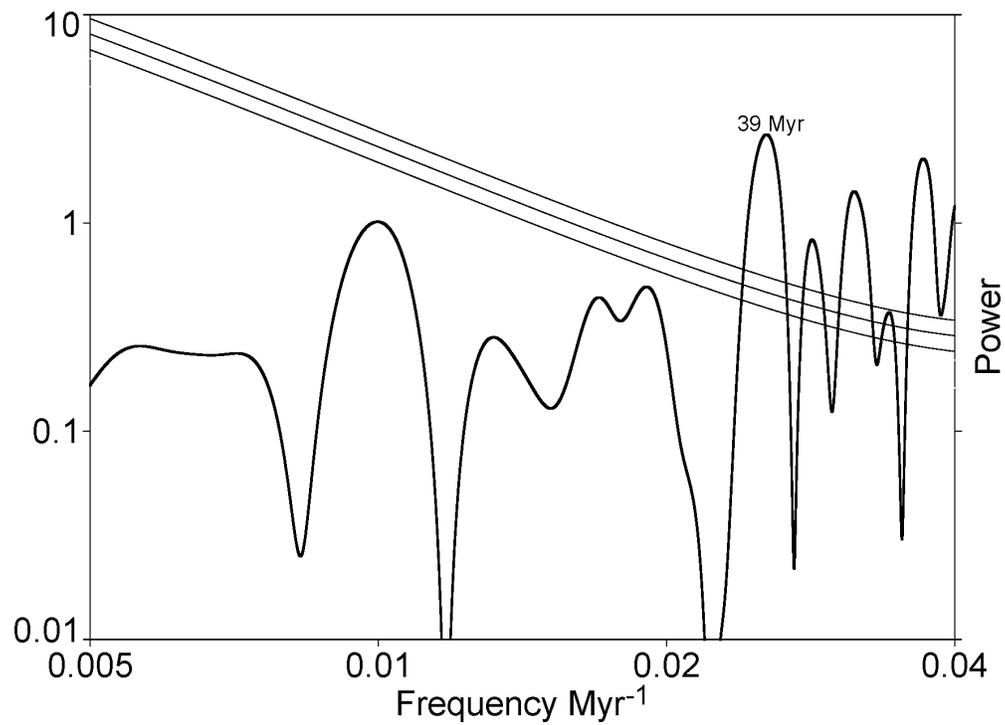

FIGURE 5.  A logarithmic plot of the power spectrum of interval lengths in the PBDB data (Alroy

et al. 2008).  The function studied is the *length* of the intervals as a function of their midpoint.

This shows a strong peak at a period of 39 Myr, which, as argued in the text, may contribute to a splitting of the 62-Myr biodiversity component into two bands, as seen in Figure 4B,C.

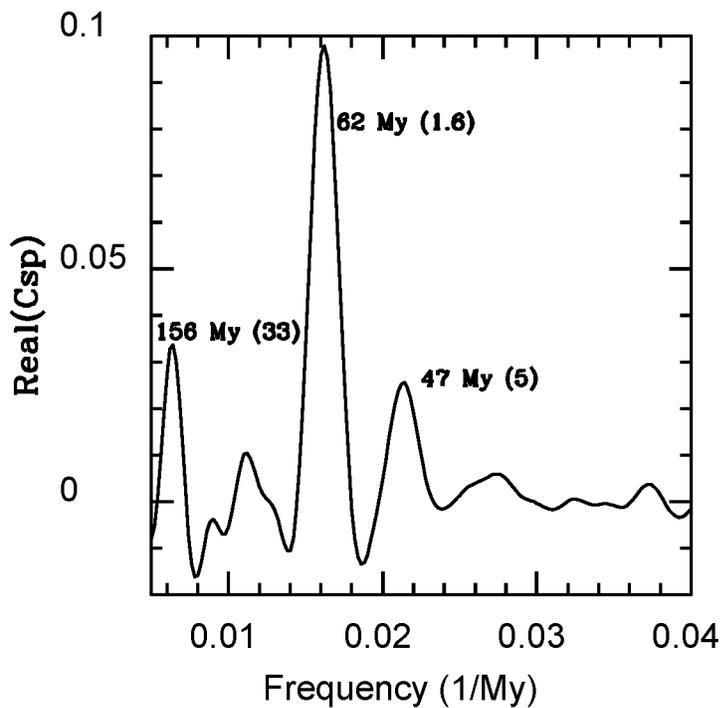

FIGURE 6.   The real part of the complex cross-spectrum of the Sepkoski/R&M data with the PBDB data, (each divided by its standard deviation in order to put them on an equal footing). This is a measure of the combined power of the same frequency in both data sets, with the same phase angles, so that the peaks coincide (Melott 2008). The inset figures give the period corresponding to the shown frequency peak.  The figure in parentheses is the mismatch, in Myr, between the peaks in one set versus the other.  The 62-Myr cycle dominates the figure and has excellent phase agreement between the two compendia.  Note that the power at higher frequencies $f > 0.03$ that appears in various spectra throughout this study do not survive this cross-check, nor do the others. *doi:10.1371/journal.pone.0004044.g003.*

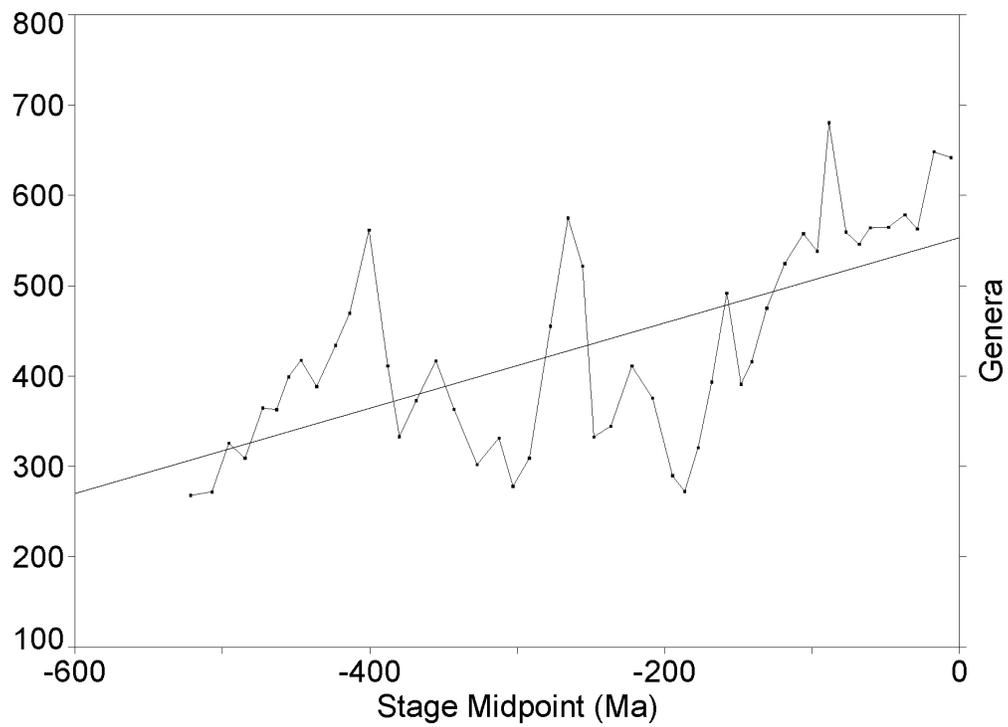

FIGURE 7. A, This straight line is the best least-squares *linear* fit to the PBDB data analyzed here.

Its coefficient of determination is $r = 0.43$.  Unlike the cubic, the straight line does not have a preferred dominant frequency within our range of interest.

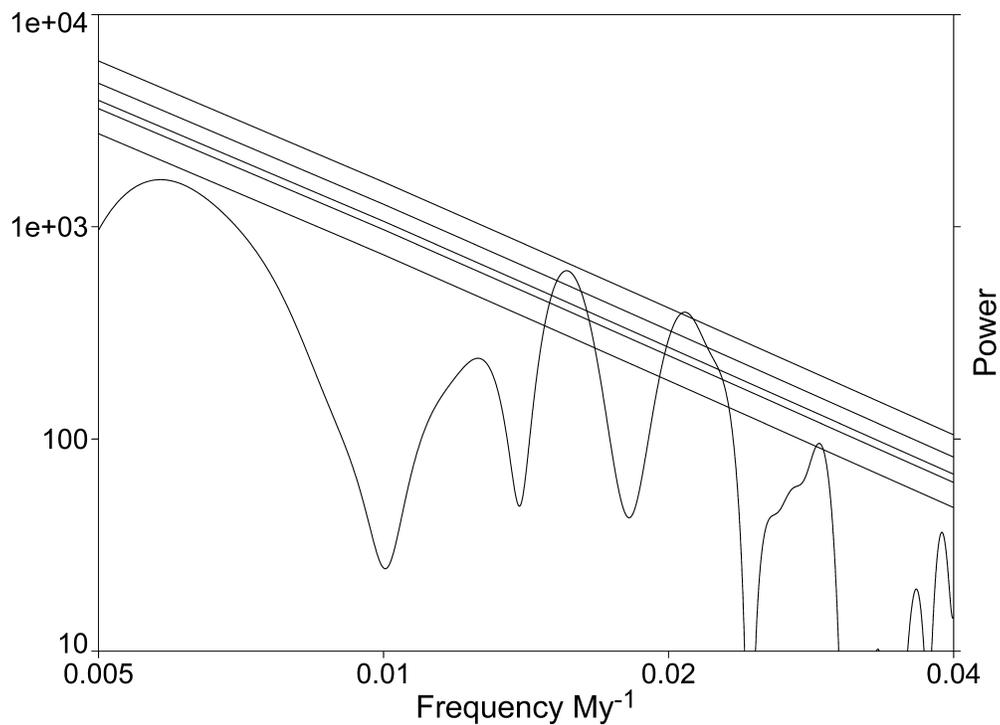

7B, The power spectrum resulting from analysis of the residuals about the straight line fit in A. Note that now the ~150 Myr peak is far below the level of significance. However, the other two features found before are still quite strong. Because peaks are lowered, we include confidence lines corresponding to $p = 0.5, 0.1, 0.05, 0.01$, and $0.001$.

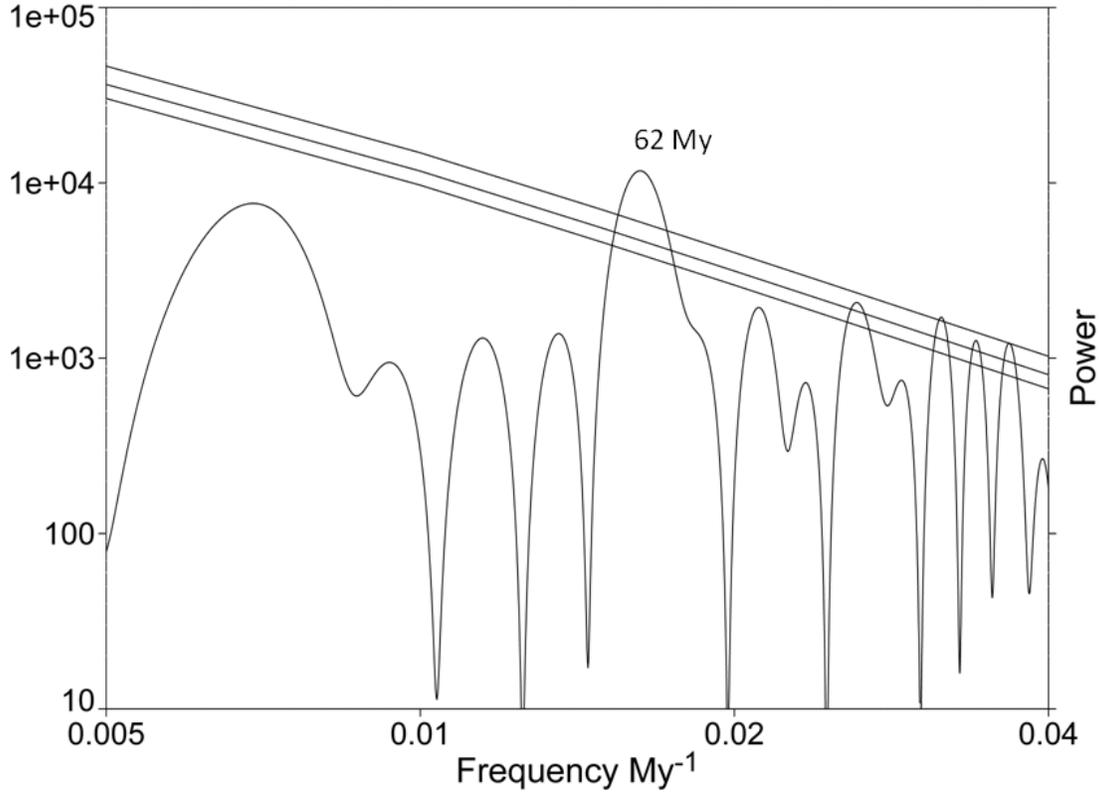

FIGURE 8. The power spectrum resulting from FFT analysis of the residuals about a cubic fit to the Sepkoski-RKB substage data. The periodicity of 62.4 ± 3 Myr has a confidence level better than 0.001, higher than found in the R&M sort of the Sepkoski data and the cross-checking papers of L&M and Cornette (2007).  Other peaks in the vicinity of 30 and 40 Myr appear here, again more strongly than in Lieberman and Melott (2007), but are not supported by the cross-checks conducted elsewhere.

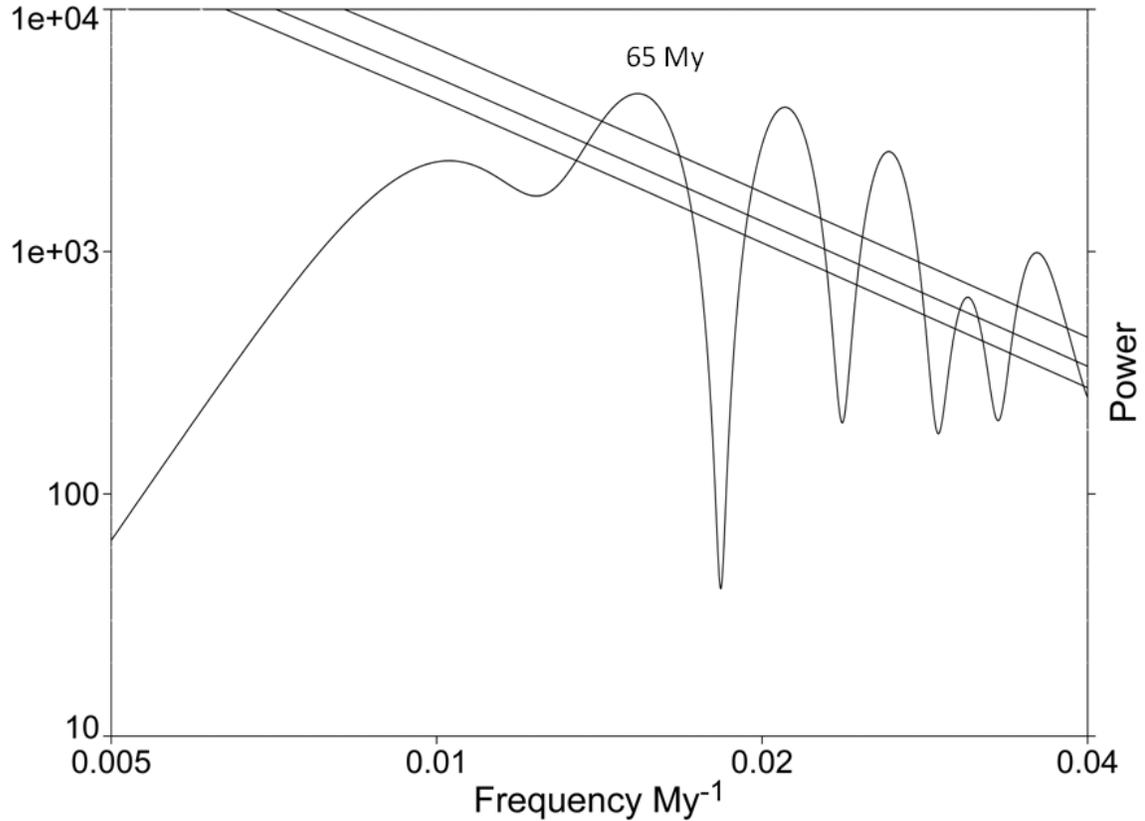

FIGURE 9. A, Power spectrum of the short-lived fauna (those which have survived less than 45 Myr) fluctuations from 45 to 295 Ma, about half the time period we examine.  Note that a peak appears at 65 Myr, although a variety of other fluctuations have arisen due to the short sample time.  Still the 65 Myr peak is consistent with previous appearances.

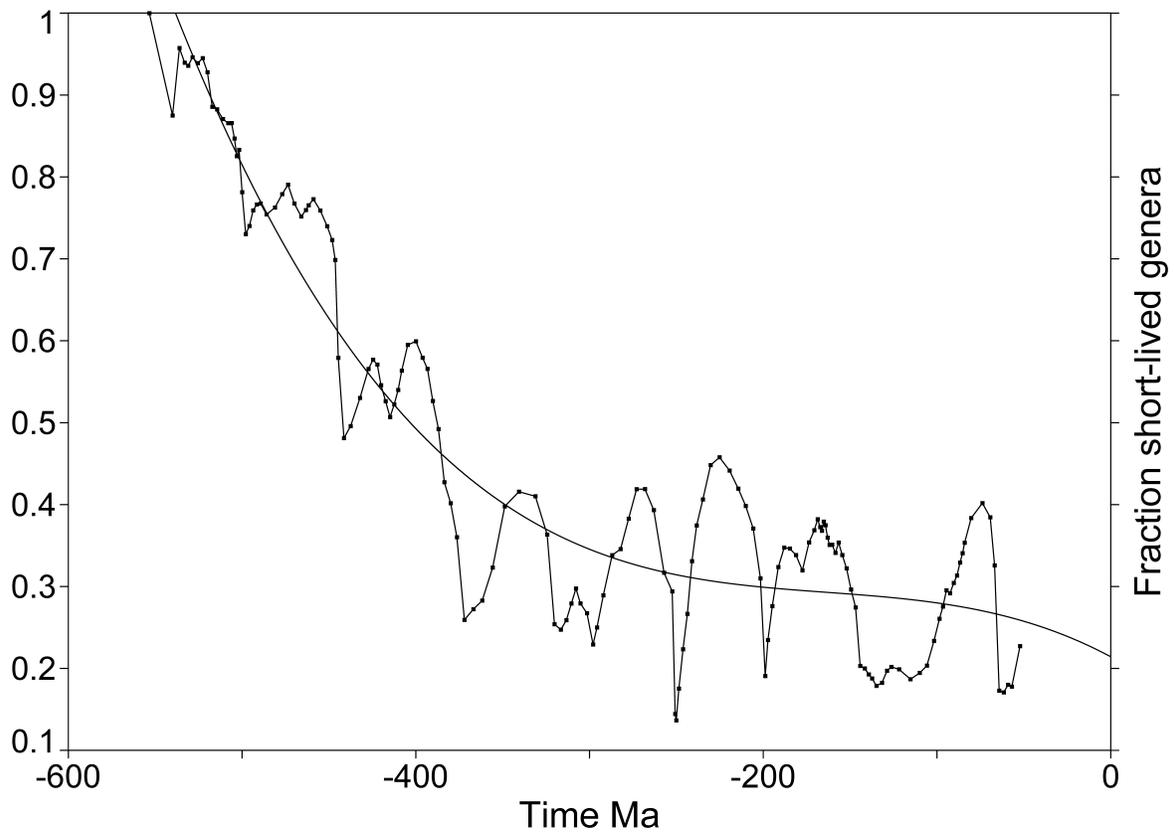

9B, In the R&M sample, the fraction of the total that belongs to the short-lived group, i.e., those that survived less than 45 Myr. The function declines due to the accumulation of survivors, and the fading of the 62-Myr cycle can be explained by lowered portion of short-lived fauna, which express the signal most strongly.

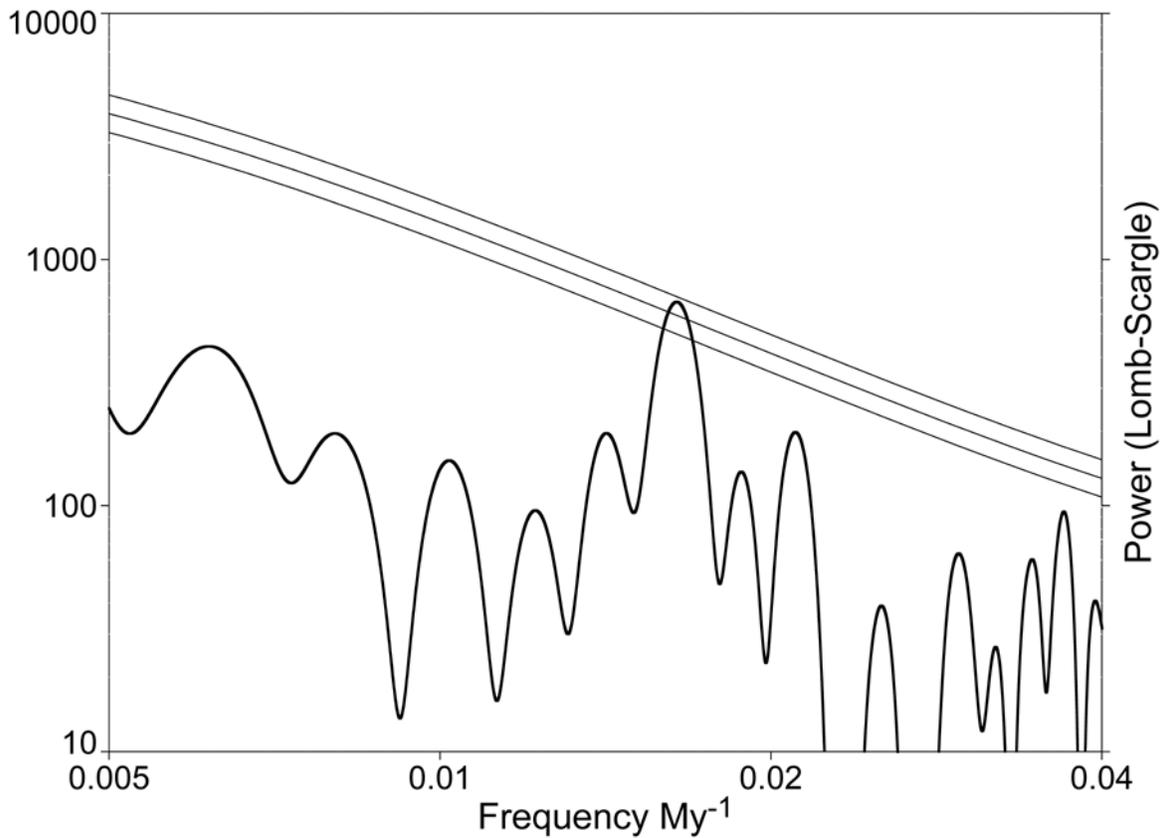

FIGURE 10. A, The fluctuation power spectrum of the number of marine families from the Fossil Record 2 summary, after detrending by a cubic fit, determined by Fast Fourier Transform. Significance levels correspond to $p = 0.05$, 0.01, and 0.001. The strongest peak corresponds to a period of about 61 Myr.

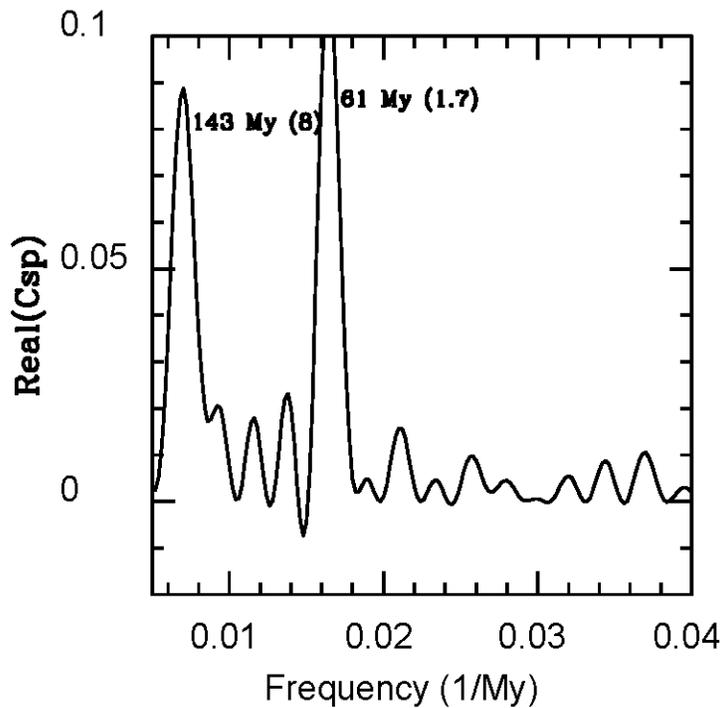

10B, The real part of the cross-spectrum between the FR2 and R&M data. The smaller peak at 143 Myr implies that, although independently significant peaks do not exist in the two compendia, the existing power is somewhat in phase. The peak at 61 Myr shows that the peaks which are significant in each data set independently are also in phase with one another. Note that the power at higher frequencies $f > 0.03$ that appears in various spectra throughout this study does not survive this cross-check, nor in Figure 6 nor 10c.

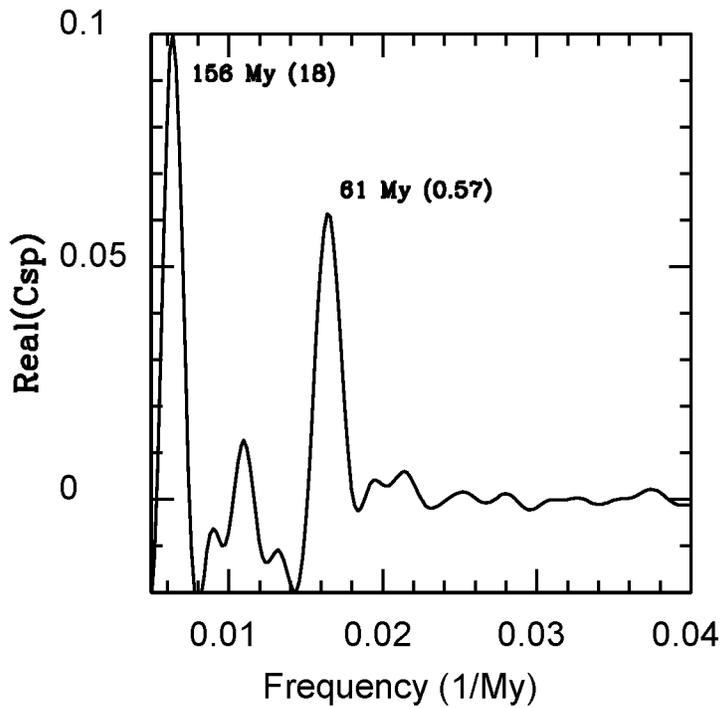

10C, The real part of the cross-spectrum between the FR2 and PBDB data.  Again, there is a peak at 61 Myr with excellent phase agreement, and another at 156 Myr has fairly poor phase agreement.  Once again power at high frequencies does not seem to have any commonality between the samples, so only one peak survives as in unambiguous agreement between all the data sets.  The range of *y*-values will change in all the cross-spectra that follow.

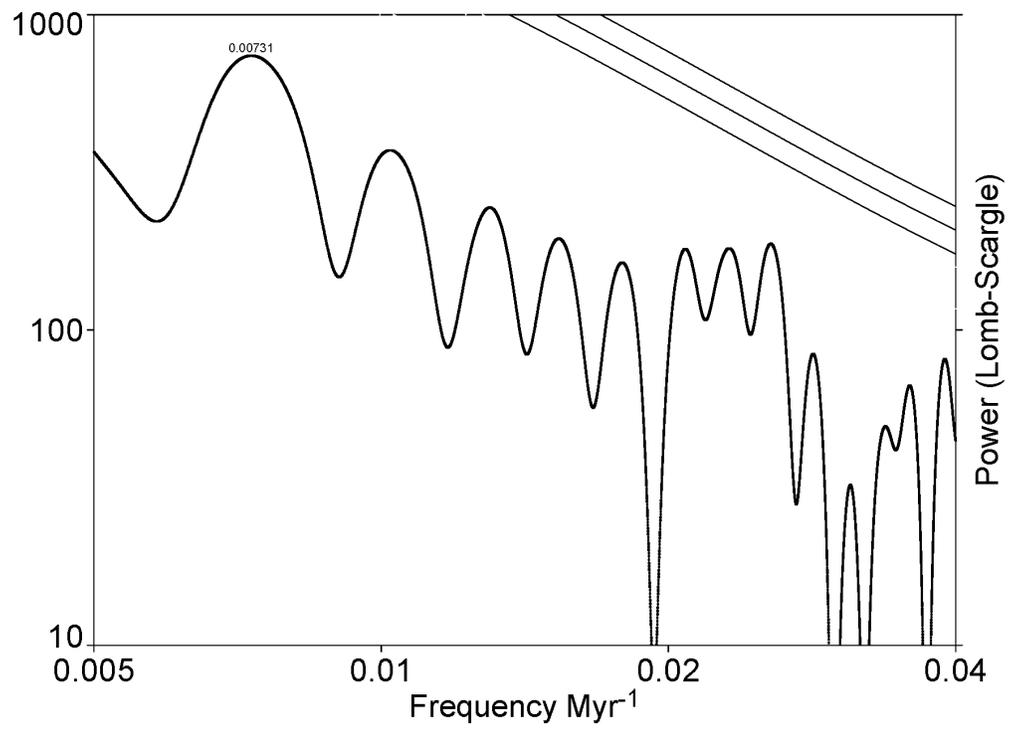

10D, The power spectrum of Fossil Record 2 continental families, "minimum" option. There are no significant peaks.